\documentclass[english]{article}
\usepackage[T1]{fontenc}
\usepackage[latin9]{inputenc}
\usepackage[letterpaper]{geometry}
\usepackage{babel}
\usepackage{units}
\usepackage{mathrsfs}
\usepackage{enumitem}
\usepackage{amsmath}
\usepackage{amssymb}
\usepackage{wasysym}
\usepackage[unicode=true]
 {hyperref}

\makeatletter

\usepackage{cite}
\usepackage{tensind}
\tensordelimiter{?}
\usepackage{bm}
\usepackage{tocloft}

\makeatother

\begin{document}

\title{Generalized Nordström Theory Revisited Part II:\\
Nordström \& Maxwell United}

\author{Johan Bengtsson\\
\\
Member of Center for Accelerator Science and Education\\
(Stony Brook University and Brookhaven National Laboratory)}

\date{\today}
\maketitle
\begin{center}
\vspace{1in}
\par\end{center}
\begin{abstract}
\noindent In 1945 Einstein concluded that \cite{Meaning_of_Rel.}:
\textquotedblleft \textit{The present theory of relativity is based
on a division of physical reality into a metric field (gravitation)
on the one hand, and into an electromagnetic field and matter on the
other hand. In reality space will probably be of a uniform character
and the present theory be valid only as a limiting case. For large
densities of field and of matter, the field equations and even the
field variables which enter into them will have no real significance.}\textquotedblright .
The dichotomy is resolved by introducing a complex Randers metric
with a real valued scalar field and complex valued vector field, providing
a unified mathematical framework for gravitation \& electromagnetism
for which the resulting theory's predictions agree with General Relativity;
to leading order in the gravitational constant. Hence, the related
experimental results validate both theories; and the former theory's
metric solutions are free of spurious singularities, because its stress-energy
tensor includes the energy \& momentum for the gravitational field;
like e.g. Maxwell's stress-energy tensor contains the electromagnetic
field.\addcontentsline{toc}{section}{\textbf{Abstract}}

\newpage{}

\tableofcontents{}\newpage{}
\end{abstract}

\section{Introduction}

\subsection{Outline}

The outline of the paper is:
\begin{enumerate}
\item The exponential metric introduced in the first part, ref. \cite{Bengtsson},
is generalized to a complex Randers metric.
\item Matter is introduced by a real valued matter and complex valued charge
density.
\item The field equations for the gravitational and electromagnetic fields
are obtained by Geometric Calculus.
\item The equations of motion for point particles are obtained by Lagrangian
mechanics.
\item The corresponing known metric solutions for a:
\begin{itemize}
\item spherically symmetric body (Schwarzschild),
\item static charged spherically symmetric body (Reissner-Nordström),
\item rotating uncharged axisymmetric body (Kerr),
\item rotating charged axisymmetric body (Kerr-Newman)
\end{itemize}
are obtained; which are free of spurious singularities.
\end{enumerate}

\section{Gravitation \& Electromagnetism: Mathematical Framework}

\subsection{Complex Randers Metric and Lagrangian Density}

Already 1918 Weyl attempted to include electromagnetism to General
Relativity (GR) by generalizing to the conformal group \cite{Weyl_1,Weyl_2};
but Einstein pointed out a fundamental flaw (sharpness of spectral
lines) for the resulting classical theory (non-integrable gravitational
scale factor) \cite{Einstein_1918}. It is avoided in quantum mechanics
(wave-mechanical phase factor) \cite{London}. Randers 1941 obtained
a geodesic with the Lorentz force from the metric and action \cite{Randers}

\begin{equation}
ds=\sqrt{g_{rs}dx^{r}dx^{s}}+A_{r}dx^{r},\qquad S=\int ds.
\end{equation}
Soh 1933 made an attempt with a complex metric \cite{Soh}. Similarly,
Einstein 1945 also considered a complex metric in his quest for a
unified field theory \cite{Einstein_Complex_1,Einstein_Complex_2,Einstein_Complex_3}.

Starting from the modified theory of General Relativity outlined in
the first part, ref. \cite{Bengtsson}, electromagnetism can be included
by introducing:
\begin{enumerate}
\item A complex Randers metric\cite{Randers}
\begin{equation}
ds=\sqrt{-g_{rs}dx^{r}dx^{s}}+i\sqrt{\frac{\kappa}{2\mu_{0}}}A_{r}dx^{r}=\left(\sqrt{-u_{r}u^{r}}+i\sqrt{\frac{\kappa}{2\mu_{0}}}u_{r}A^{r}\right)d\tau\label{eq:Line_Element}
\end{equation}
where $A_{j}\left(x\right)=\left[\phi_{\mathrm{q}},\bar{A}\right]$
is the electromagnetic 4-potential, with the norm
\begin{align}
ds^{2} & =-dsds^{\ast}=\left[u_{r}u^{r}-\frac{\kappa}{2\mu_{0}}\left(u_{r}A^{r}\right)^{2}\right]d\tau^{2}=g_{rs}dx^{r}dx^{s}-\frac{\kappa\varepsilon_{0}}{2}\left(v_{r}A^{r}\right)^{2}c_{0}^{2}dt^{2}\label{eq:Metric_UFT_1}
\end{align}
where
\begin{equation}
u^{i}\equiv\frac{dx^{i}}{d\tau},\qquad u_{r}u^{r}=-c_{0}^{2},\qquad\kappa\equiv\frac{8\pi G}{c_{0}^{4}},\qquad c_{0}=\frac{1}{\sqrt{\mu_{0}\varepsilon_{0}}}
\end{equation}
 and $g_{ij}$ the metric tensor
\begin{align}
g_{rs}dx^{r}dx^{s} & =-\frac{c_{0}^{2}dt^{2}}{\gamma_{\phi_{\mathrm{g}}}^{2}}=-e^{\nicefrac{2\phi_{\mathrm{g}}}{c_{0}^{2}}}\left[1-\left(\frac{v}{c_{0}e^{\nicefrac{2\phi_{\mathrm{g}}}{c_{0}^{2}}}}\right)^{2}\right]c_{0}^{2}dt^{2}\nonumber \\
 & =-e^{\nicefrac{2\phi_{\mathrm{g}}}{c_{0}^{2}}}c_{0}^{2}dt^{2}+e^{-\nicefrac{2\phi_{\mathrm{g}}}{c_{0}^{2}}}\left(dx^{2}+dy^{2}+dz^{2}\right)\nonumber \\
 & =-e^{\nicefrac{2\phi_{\mathrm{g}}}{c_{0}^{2}}}c_{0}^{2}dt^{2}+e^{-\nicefrac{2\phi_{\mathrm{g}}}{c_{0}^{2}}}\left(dr^{2}+r^{2}d\theta^{2}+r^{2}\sin^{2}\left(\theta\right)d\varphi^{2}\right)\nonumber \\
 & =e^{-\nicefrac{2\phi_{\mathrm{g}}}{c_{0}^{2}}}\left(-e^{\nicefrac{4\phi_{\mathrm{g}}}{c_{0}^{2}}}c_{0}^{2}dt^{2}+dr^{2}+r^{2}d\theta^{2}+r^{2}\sin^{2}\left(\theta\right)d\varphi^{2}\right)\label{eq:Metric_UFT_2}
\end{align}
with the determinant (Cartesian coordinates)
\begin{equation}
\sqrt{-g}=e^{-\nicefrac{2\phi_{\mathrm{g}}}{c_{0}^{2}}}\label{eq:Det_UFT}
\end{equation}
where
\begin{equation}
\gamma_{\phi_{\mathrm{g}}}\equiv\frac{1}{e^{\nicefrac{2\phi_{\mathrm{g}}}{c_{0}^{2}}}\sqrt{1-\left(\frac{v}{c_{0}e^{\nicefrac{2\phi_{\mathrm{g}}}{c_{0}^{2}}}}\right)^{2}}}=\frac{1}{e^{\nicefrac{2\phi_{\mathrm{g}}}{c_{0}^{2}}}\sqrt{1-\beta_{\phi}^{2}}},\qquad\beta_{\phi_{\mathrm{g}}}\equiv\frac{v}{c_{0}e^{\nicefrac{2\phi_{\mathrm{g}}}{c_{0}^{2}}}}.\label{eq:gamma_UFT}
\end{equation}
The origin of the conformal factor $e^{-\nicefrac{2\phi_{\mathrm{g}}}{c_{0}^{2}}}$
and conformal symmetry breaking factor $c_{0}\rightarrow c_{0}e^{\nicefrac{2\phi_{\mathrm{g}}}{c_{0}^{2}}}$
is provided in ref. \cite{Bengtsson}.
\item Matter by a real valued mass and complex valued charge density
\begin{equation}
\varepsilon_{\mathrm{m}}=\sqrt{-g}\left(\rho_{\mathrm{m}}c_{0}+i\sqrt{\frac{2\mu_{0}}{\kappa}}\rho_{\mathrm{q}}\right)\label{eq:Cmplx_Charge_Density}
\end{equation}
\item The Lagrangian density and action
\begin{equation}
\mathscr{L}=\sqrt{-g}\left(-\rho_{\mathrm{m}}c_{0}\sqrt{-u_{r}u^{r}}+\rho_{\mathrm{q}}v_{r}A^{r}\right),\qquad S=-\mathrm{Re}\left\{ \int\varepsilon_{\mathrm{m}}ds\right\} =\int\mathscr{L}d^{4}x.\label{eq:Lagr_Density_UFT}
\end{equation}
\end{enumerate}
For a gravitational point source
\begin{equation}
\phi_{\mathrm{g}}=-\frac{GM}{r}
\end{equation}
and no electromagnetic field, $A^{i}=0$, the exponential metric Eq.
(\ref{eq:Metric_UFT_2}) satisfies the algebraic relation \cite{Bengtsson}
(Einstein's equations \cite{Einstein_GR})
\begin{equation}
?R^{i}{}_{j}?-\frac{1}{2}\delta_{j}^{i}=\kappa?T^{i}{}_{j}?
\end{equation}
with (Riemann tensor)
\begin{equation}
?R^{i}{}_{j}?=-\frac{2}{c_{0}^{4}}\phi^{i}\phi_{j}
\end{equation}
and (stress-energy tensor density)
\begin{equation}
?{\mathscr{T}}^{i}{}_{j}?=\sqrt{-g}?T^{i}{}_{j}?=-\frac{\sqrt{-g}}{4\pi G}\left(\phi^{i}\phi_{j}-\frac{1}{2}\delta_{j}^{i}\phi^{r}\phi_{r}\right)=\frac{GM^{2}}{8\pi}\left[\begin{array}{cccc}
\frac{1}{r^{4}} & 0 & 0 & 0\\
0 & -\frac{1}{r^{4}} & 0 & 0\\
0 & 0 & \frac{1}{r^{4}} & 0\\
0 & 0 & 0 & \frac{1}{r^{4}}
\end{array}\right],\qquad?{\mathscr{T}}^{r}{}_{r}?=\frac{GM^{2}}{4\pi r^{4}}
\end{equation}
which includes the gravitational field; whereas for GR (vacuum)

\begin{eqnarray}
?{\mathscr{T}}^{i}{}_{j}? & = & 0.
\end{eqnarray}
Hence, not only the covariant but also the ordinary divegence is zero

\begin{equation}
?{\mathscr{T}}^{ri}{}_{;r}?=0,\qquad?{\mathscr{T}}^{ri}{}_{,r}?=0.
\end{equation}
and the energy-momentum for the gravitational field and sources are
conserved. In other words, like for an electrostatic point charge
\begin{equation}
\phi_{\mathrm{q}}=\frac{Q}{4\pi\varepsilon_{0}r}
\end{equation}
and the Maxwell stress-energy tensor for Minkowski metric
\begin{equation}
?T^{i}{}_{j}?=\frac{1}{\mu_{0}}\left(F^{ir}F_{jr}-\frac{1}{4}\delta_{j}^{i}F^{rs}F_{rs}\right)=\frac{Q^{2}}{32\pi^{2}\varepsilon_{0}}\begin{bmatrix}-\frac{1}{r^{4}} & 0 & 0 & 0\\
0 & -\frac{1}{r^{4}} & 0 & 0\\
0 & 0 & \frac{1}{r^{4}} & 0\\
0 & 0 & 0 & \frac{1}{r^{4}}
\end{bmatrix},\qquad?T^{r}{}_{r}?=0.
\end{equation}

\subsection{Field Equations: Geometric Calculus}

Based on Grassmann's exterior product \cite{Grassmann}
\begin{equation}
a\land b=-b\land a.
\end{equation}
Cartan 1899 generalized vector calculus by introducing exterior calculus
\cite{Cartan}. The exterior derivative of a smooth function $f$
is the differential
\begin{equation}
df
\end{equation}
such that
\begin{equation}
d\left(df\right)=0.
\end{equation}
More generally
\begin{equation}
d\left(\alpha\wedge\beta\right)=d\alpha\wedge\beta+\left(-1\right)^{p}\alpha\wedge d\beta
\end{equation}
and
\begin{equation}
d\circ d=0
\end{equation}
where $\alpha$ is a $p$-form.

In local coordinates, the exterior derivative of a simple $k$-form
\begin{equation}
\varphi=adx^{I}=adx^{i_{1}}\wedge dx^{i_{2}}\ldots\wedge dx^{i_{k}}
\end{equation}
is
\begin{equation}
d\varphi\equiv dx^{i}\wedge\partial_{i}\varphi=\partial_{i}adx^{i}\wedge dx^{I}
\end{equation}
and for a general $k$-form
\begin{equation}
\omega=\omega_{I}dx^{I},\qquad d\omega=dx^{i}\wedge\partial_{i}\omega_{I}=\partial_{i}\omega_{I}dx^{i}\wedge dx^{I}.
\end{equation}
For example, for a 1-form $\omega$
\begin{equation}
\left(d\omega\right)_{ij}=\partial_{i}\omega_{j}-\partial_{j}\omega_{i}.
\end{equation}
Hence, the framework provides for a concise summary of Maxwell's equations
\cite{Cartan_Maxwell} 
\begin{equation}
\begin{array}{c}
\begin{array}{ccccc}
\begin{array}{c}
\mathrm{gauge}\\
\varphi
\end{array} & \rightarrow & \begin{array}{c}
\mathrm{vector\:potential}\\
A=d\varphi
\end{array} & \rightarrow\begin{array}{c}
\mathrm{field\:tensor}\\
F=dA
\end{array} & \rightarrow\begin{array}{c}
f\mathrm{ield\:equations}\\
dF=0
\end{array}\end{array}\end{array}.
\end{equation}
Similarly, Geometric Algebra (GA) can be utilized to generate the
field equations for the complex Randers metric. Starting from the
Geometric Product \cite{Grassmann,Hamilton,Peano,Hestenes,Hestenes_2,Doran,Vold}
\begin{equation}
AB\equiv A\cdot B+A\land B
\end{equation}
calculus is introduced by
\begin{equation}
\nabla A\equiv\nabla\cdot A+\nabla\wedge A
\end{equation}
with
\begin{equation}
\nabla\cdot A_{k}\equiv\left\langle \nabla A_{k}\right\rangle _{k-1}=e^{r}\cdot\partial_{r}A_{k},\qquad\nabla\wedge A_{k}\equiv\left\langle \nabla A_{k}\right\rangle _{k+1}=e^{r}\wedge\partial_{r}A_{k}
\end{equation}
for an $k$-grade multivector field. It follows that
\begin{equation}
\nabla^{2}A=\nabla\left(\nabla A\right)=\nabla\cdot\left(\nabla\wedge A\right)+\nabla\land\left(\nabla\cdot A\right)
\end{equation}
since
\begin{equation}
\nabla\wedge\left(\nabla\phi\right)=0,\qquad\nabla\cdot\left(\nabla\cdot A\right)=0,\qquad\nabla\wedge\left(\nabla\wedge A\right)=0.
\end{equation}
Space-Time Algebra (STA) is generated by introducing the four vectors
$\left\{ \gamma_{i}\right\} $ satisfying
\begin{equation}
\gamma_{i}\cdot\gamma_{j}=\frac{1}{2}\left(\gamma_{i}\gamma_{j}+\gamma_{j}\gamma_{i}\right)=\eta_{ij}
\end{equation}
with (pseudoscalar)
\begin{equation}
I\equiv\gamma_{0}\gamma_{1}\gamma_{2}\gamma_{3}.
\end{equation}
Hence, the algebraic properties of STA are those of the Dirac matrices.

In the rest frame defined by the $\gamma_{0}$ vector one introduces
\begin{equation}
\sigma_{i}\equiv\gamma_{i}\gamma_{0},\qquad\frac{1}{2}\left(\sigma_{i}\sigma_{j}+\sigma_{j}\sigma_{i}\right)=\delta_{ij},\qquad\sigma_{i}\sigma_{i}\sigma_{i}=I
\end{equation}
i.e., the algebra properties of space are those of the Pauli matrices;
the even subalgebra of STA.

In Euclidian 3-space for $A,B$ 1-grade (vector valued functions),
the mathematical framework simplifies to the traditional vector calculus
(introduced by Gibbs \cite{Gibbs}; vs. Hamilton, Grassman, Peano,
etc. \cite{Grassmann,Hamilton,Peano})
\begin{equation}
A\land B=IA\times B,\qquad\bar{\nabla}\wedge A=I\bar{\nabla}\times A
\end{equation}
with
\begin{equation}
\nabla^{2}=\text{\Square},\qquad\gamma_{0}\cdot\nabla=\frac{1}{c_{0}}\partial_{t},\qquad\gamma_{0}\land\nabla=\bar{\nabla}.
\end{equation}
The line element is
\begin{align}
ds & =\sqrt{-g_{rs}dx^{r}dx^{s}}+i\sqrt{\frac{\kappa}{2\mu_{0}}}A_{r}dx^{r}=\frac{1}{\gamma_{\phi}}c_{0}dt+i\sqrt{\frac{\kappa}{2\mu_{0}}}A_{r}dx^{r}\nonumber \\
 & =\sqrt{e^{\nicefrac{2\phi_{\mathrm{g}}}{c_{0}^{2}}}-e^{-\nicefrac{2\phi_{\mathrm{g}}}{c_{0}^{2}}}\left(\frac{v}{c_{0}}\right)^{2}}c_{0}dt+i\sqrt{\frac{\kappa}{2\mu_{0}}}A_{r}dx^{r}\nonumber \\
 & =\left\{ 1-\frac{1}{2}\left(\frac{v}{c_{0}}\right)^{2}+\left[1+\left(\frac{v}{c_{0}}\right)^{2}\right]\frac{\phi_{\mathrm{g}}}{c_{0}^{2}}\right\} c_{0}dt+i\sqrt{\frac{\kappa}{2\mu_{0}}}A+\ldots,
\end{align}
where $A$ is a 1-form. Prolonging twice gives
\begin{align}
\partial ds & =\left[1+\left(\frac{v}{c_{0}}\right)^{2}\right]\frac{\nabla\phi_{\mathrm{g}}}{c_{0}^{2}}c_{0}dt+i\sqrt{\frac{\kappa}{2\mu_{0}}}\left(\nabla\cdot A+\nabla\wedge A\right)\nonumber \\
 & =\left[1+\left(\frac{v}{c_{0}}\right)^{2}\right]\frac{\nabla\phi_{\mathrm{g}}}{c_{0}^{2}}c_{0}dt+i\sqrt{\frac{\kappa}{2\mu_{0}}}\left(\nabla\cdot A+F\right),\nonumber \\
\partial^{2}ds & =\left[1+\left(\frac{v}{c_{0}}\right)^{2}\right]\frac{\square\phi_{\mathrm{g}}}{c_{0}^{2}}c_{0}dt+i\sqrt{\frac{\kappa}{2\mu_{0}}}\left[\nabla\left(\nabla\cdot A\right)+\nabla F\right]
\end{align}
and setting the R.H.S. to zero one obtains
\begin{equation}
\square\phi_{\mathrm{g}}=0,\qquad\nabla\left(\nabla\cdot A\right)+\nabla F=0
\end{equation}
i.e., the wave equation for the gravitational field and Maxwell's
equations for vacuum. For the Lorentz gauge
\begin{equation}
\nabla\cdot A=0.
\end{equation}
By introducing the electromagnetic bivector (Riemann\textendash Silberstein
vector \cite{Weber,Silberstein})
\begin{equation}
F\equiv\nabla\wedge A=\sqrt{\varepsilon_{0}}\left(\bar{E}+Ic_{0}\bar{B}\right)=\sqrt{\varepsilon_{0}}\left(E^{r}\sigma_{r}+Ic_{0}B^{r}\sigma_{r}\right),
\end{equation}
and contrarily
\begin{equation}
\sqrt{\varepsilon_{0}}\bar{E}=\frac{1}{2}\left(F-\gamma_{0}F\gamma_{0}\right),\qquad Ic_{0}\bar{B}=\frac{1}{2}\left(F+\gamma_{0}F\gamma_{0}\right),
\end{equation}
the Maxwell stress-energy tensor is
\begin{equation}
T=FF=F\cdot F+F\land F.
\end{equation}
Similarly, the energy density and Poynting vector are 
\begin{equation}
\frac{1}{2}\left|F\right|^{2}=\frac{1}{2}FF^{\dagger}=\left[\frac{1}{2}\left(\varepsilon_{0}E^{2}+\frac{1}{\mu_{0}}B^{2}\right)+I\frac{1}{\mu_{0}c_{0}}\bar{E}\land\bar{B}\right]=\left[\frac{1}{2}\left(\varepsilon_{0}E^{2}+\frac{1}{\mu_{0}}B^{2}\right)-\frac{1}{\mu_{0}c_{0}}\bar{E}\times\bar{B}\right].
\end{equation}

\subsection{Equations of Motion for Point Particles: Lagrangian Mechanics}

The Lagrangian density Eq. (\ref{eq:Lagr_Density_UFT}) for a point
particle is
\begin{equation}
\mathscr{L}=\left(-\rho_{\mathrm{m}}c_{0}\sqrt{-u_{r}u^{r}}+\rho_{\mathrm{q}}u^{r}A_{r}\right)\delta^{3}\left(x\right)
\end{equation}
with the Lagrangian and action
\begin{equation}
L=\int\mathscr{L}d^{3}x=-m_{0}c_{0}\sqrt{-u_{r}u^{r}}+qu^{r}A_{r}=-m_{0}c_{0}\sqrt{-g_{rs}u^{r}u^{s}}+qg_{rs}u^{r}A^{s},\qquad S=\int Ld\tau.\label{eq:Lagrangian_UFT}
\end{equation}
 The equations of motion are (covariant Euler-Lagrange equations)
\begin{equation}
\frac{d}{d\tau}\left(\partial_{u^{i}}L\right)-\partial_{i}L=0
\end{equation}
which leads to
\begin{align}
\frac{d}{d\tau}\left(\partial_{u^{i}}L\right) & =\frac{d}{d\tau}\left(m_{0}g_{ir}u^{r}+qA_{i}\right)=m_{0}\left(g_{ir}\frac{du^{r}}{d\tau}+u^{r}\frac{dg_{ir}}{d\tau}\right)+qu^{r}\partial_{r}A_{i}\nonumber \\
 & =m_{0}\left(g_{ir}\frac{du^{r}}{d\tau}+u^{r}u^{s}\partial_{s}g_{ir}\right)+qu^{r}\partial_{r}A_{i}\nonumber \\
 & =m_{0}\left[g_{ir}\frac{du^{r}}{d\tau}+\frac{1}{2}u^{r}u^{s}\left(\partial_{s}g_{ir}+\partial_{r}g_{si}\right)\right]+qu^{r}\partial_{r}A_{i},\nonumber \\
-\partial_{i}L & =-m_{0}u^{r}u^{s}\frac{1}{2}\partial_{i}g_{rs}-qu^{r}\partial_{i}A_{r}\label{eq:Euler_Lagrange_cov}
\end{align}
where we have used
\begin{equation}
u_{r}u^{r}=-c_{0}^{2}.
\end{equation}
It follows that
\begin{align}
\frac{d}{d\tau}\left(\partial_{u^{i}}L\right)-\partial_{i}L & =m_{0}g_{ir}\frac{du^{r}}{d\tau}+m_{0}u^{r}u^{s}\frac{1}{2}\left(\partial_{s}g_{ir}+\partial_{r}g_{si}-\partial_{i}g_{rs}\right)-qu^{r}F_{ir}=0
\end{align}
where (electromagnetic field tensor)
\begin{equation}
F_{ij}\equiv\partial_{i}A_{j}-\partial_{j}A_{i}=A_{j;i}-A_{i;j}
\end{equation}
and by raising the index $i$, it can be written (geodesic with Lorentz
force)

\begin{equation}
\frac{du^{i}}{d\tau}+?\Gamma^{i}{}_{rs}?u^{r}u^{s}=\frac{q}{m_{0}}u^{r}?F^{i}{}_{r}?\label{eq:Eqs-of-Motion}
\end{equation}
where (Christoffel symbols \cite{Christoffel})
\begin{equation}
?\Gamma^{i}{}_{jk}?=\frac{1}{2}g^{ir}\left(\partial_{k}g_{rj}+\partial_{j}g_{rk}-\partial_{r}g_{jk}\right).
\end{equation}
The conjugate momentum is introduced by
\begin{equation}
p_{i}\equiv\partial_{u^{i}}L=m_{0}u_{i}+qA_{i}
\end{equation}
and the Hamiltonian is generated by (Legrenge transformation)
\begin{equation}
H=u_{r}p^{r}-L=\frac{m_{0}}{2}u_{r}u^{r}=\frac{1}{2m_{0}}g_{rs}\left(p^{r}-qA^{r}\right)\left(p^{s}-qA^{s}\right)\label{eq:Cov_Ham}
\end{equation}
and the equations of motion are (Hamiltons equations)
\begin{equation}
\frac{dx^{i}}{d\tau}=\partial_{p_{i}}H,\qquad\frac{dp_{i}}{d\tau}=-\partial_{i}H.
\end{equation}
In conclusion, the outlined mathematical framework generates the equations
of motion for GR with the Lorentz force Eq. (\ref{eq:Eqs-of-Motion})
and the corresponding covariant Hamiltonian Eq. (\ref{eq:Cov_Ham}).

\section{Solutions}

Since the metric Eqs. (\ref{eq:Metric_UFT_1}) and (\ref{eq:Metric_UFT_2})
is a functional for the gravitational and electromagnetic potentials,
solving for the latter provides the metric solution directly. Besides,
after a static solution has been found, related dynamic solutions
can be found by coordinate transformations; e.g. to a rotating co-moving
frame.

\subsection{Static Spherically Symmetric Body}

For a spherically symmetric body with mass $M$
\begin{equation}
\phi_{\mathrm{g}}=-\frac{GM}{r}
\end{equation}
the exponential metric Eq. (\ref{eq:Metric_UFT_2}) is

\begin{align}
ds^{2} & =-e^{-\nicefrac{r_{\mathrm{S}}}{r}}c_{0}^{2}dt^{2}+e^{\nicefrac{r_{\mathrm{S}}}{r}}\left(dr^{2}+r^{2}d\theta^{2}+r^{2}\sin^{2}\left(\theta\right)d\varphi^{2}\right)\nonumber \\
 & =-\left(1-\frac{r_{\mathrm{S}}}{r}+\ldots\right)c_{0}^{2}dt^{2}+\left(1+\frac{r_{\mathrm{S}}}{r}+\ldots\right)\left(dr^{2}+r^{2}d\theta^{2}+r^{2}\sin^{2}\left(\theta\right)d\varphi^{2}\right)
\end{align}
where
\begin{equation}
r_{\mathrm{S}}\equiv\frac{2GM}{c_{0}^{2}}.
\end{equation}
The Schwarzschild metric for GR in isotropic form is \cite{Eddington}
\begin{align}
ds^{2} & =-\frac{\left(1-\frac{r_{\mathrm{S}}}{4r}\right)^{2}}{\left(1+\frac{r_{\mathrm{S}}}{4r}\right)^{2}}c_{0}^{2}dt^{2}+\left(1+\frac{r_{\mathrm{S}}}{4r}\right)^{4}\left(dr^{2}+r^{2}d\theta^{2}+r^{2}\sin^{2}\left(\theta\right)d\varphi^{2}\right)\nonumber \\
 & =-\left(1-\frac{r_{\mathrm{S}}}{r}+\ldots\right)c_{0}^{2}dt^{2}+\left(1+\frac{r_{\mathrm{S}}}{r}+\ldots\right)\left(dr^{2}+r^{2}d\theta^{2}+r^{2}\sin^{2}\left(\theta\right)d\varphi^{2}\right)
\end{align}
which agrees to leading order.

\subsection{Static Charged Spherically Symmetric Body\label{subsec:Static_Charged_Body}}

The solutions to Poisson's equations
\begin{equation}
\triangle\phi_{\mathrm{g}}=4\pi G\rho_{\mathrm{m}},\qquad\triangle\phi_{\mathrm{q}}=-\frac{\rho_{\mathrm{q}}}{\varepsilon_{0}}
\end{equation}
for a charged spherically symmetric body with mass $M$ and charge
$Q$ are
\begin{equation}
\phi_{\mathrm{m}}=-\frac{GM}{r},\qquad\phi_{\mathrm{q}}=\frac{Q}{4\pi\varepsilon_{0}r}.
\end{equation}
The metric is obtained from Eqs. (\ref{eq:Metric_UFT_1}) and (\ref{eq:Metric_UFT_2})
which gives

\begin{align}
ds^{2} & =g_{rs}dx^{r}dx^{s}-\frac{\kappa\varepsilon_{0}}{2}\left(v_{r}A^{r}\right)^{2}c_{0}^{2}dt^{2}=-\left[\frac{1}{\gamma_{\phi_{\mathrm{g}}}^{2}}+\frac{\kappa\varepsilon_{0}}{2}\left(v_{r}A^{r}\right)^{2}\right]c_{0}^{2}dt^{2}\nonumber \\
 & =-\left(e^{-\frac{r_{\mathrm{S}}}{r}}+\frac{r_{\mathrm{Q}}^{2}}{r^{2}}\right)c_{0}^{2}dt^{2}+e^{\frac{r_{\mathrm{S}}}{r}}\left[dr^{2}+r^{2}\left(d\theta^{2}+\sin^{2}\left(\theta\right)d\varphi^{2}\right)\right]\nonumber \\
 & =-\left(1-\frac{r_{\mathrm{S}}}{r}+\frac{r_{\mathrm{Q}}^{2}}{r^{2}}\right)c_{0}^{2}dt^{2}+\left(1+\frac{r_{\mathrm{S}}}{r}\right)\left[dr^{2}+r^{2}\left(d\theta^{2}+\sin^{2}\left(\theta\right)d\varphi^{2}\right)\right]+\ldots\label{eq:Static_Charged_Body}
\end{align}
where
\begin{equation}
r_{\mathrm{S}}\equiv\frac{2GM}{c_{0}^{2}},\qquad r_{\mathrm{Q}}^{2}\equiv\frac{GQ^{2}}{4\pi\varepsilon_{0}c_{0}^{4}}.
\end{equation}
The Reissner-Nordström metric for GR is \cite{Reissner,Nordstr=0000F6m_3}
\begin{align}
ds^{2} & =-\left(1-\frac{r_{\mathrm{S}}}{r}+\frac{r_{\mathrm{Q}}^{2}}{r^{2}}\right)c_{0}^{2}dt^{2}+\frac{1}{1-\frac{r_{\mathrm{S}}}{r}+\frac{r_{\mathrm{Q}}^{2}}{r^{2}}}dr^{2}+r^{2}d\theta^{2}+r^{2}\sin^{2}\left(\theta\right)d\varphi^{2}\nonumber \\
 & =-\left(1-\frac{r_{\mathrm{S}}}{r}+\frac{r_{\mathrm{Q}}^{2}}{r^{2}}\right)c_{0}^{2}dt^{2}+\left(1+\frac{r_{\mathrm{S}}}{r}-\frac{r_{\mathrm{Q}}^{2}}{r^{2}}\right)dr^{2}+r^{2}d\theta^{2}+r^{2}\sin^{2}\left(\theta\right)d\varphi^{2}+...\label{eq:Reissner-Nordstr=0000F6m_Solution}
\end{align}
It becomes singular for
\begin{equation}
r=\frac{1}{2}\left(r_{\mathrm{S}}\pm\sqrt{r_{\mathrm{S}}^{2}-4r_{\mathrm{Q}}^{2}}\right).
\end{equation}
It can be transformed into isotropic form by a coordinate transformation
$r\rightarrow\rho$
\begin{equation}
ds^{2}=-A^{2}\left(\rho\right)c_{0}^{2}dt^{2}+B^{2}\left(\rho\right)\left[d\rho^{2}+\rho{}^{2}\left(d\theta^{2}+\sin^{2}\left(\theta\right)d\varphi^{2}\right)\right]
\end{equation}
where
\begin{equation}
A^{2}\left(\rho\right)=1-\frac{r_{\mathrm{S}}}{r}+\frac{r_{\mathrm{Q}}^{2}}{r^{2}}
\end{equation}
which gives
\begin{equation}
B^{2}\left(\rho\right)d\rho{}^{2}=\frac{dr^{2}}{1-\frac{r_{\mathrm{S}}}{r}+\frac{r_{\mathrm{Q}}^{2}}{r^{2}}},\qquad B^{2}\left(\rho\right)\rho{}^{2}=r^{2}
\end{equation}
with
\begin{equation}
\frac{d\rho}{\rho}=\frac{dr}{\sqrt{r^{2}-r_{\mathrm{S}}r+r_{\mathrm{Q}}^{2}}}
\end{equation}
and by integrating
\begin{equation}
\rho=\frac{r}{2}\left(1-\frac{r_{\mathrm{S}}}{2r}+\sqrt{1-\frac{r_{\mathrm{S}}}{r}+\frac{r_{\mathrm{Q}}^{2}}{r^{2}}}\right)
\end{equation}
and solving for $r$ 
\begin{equation}
r=\rho\left[\left(1+\frac{r_{\mathrm{S}}}{4\rho}\right)^{2}-\frac{r_{\mathrm{Q}}^{2}}{4\rho^{2}}\right].
\end{equation}
Replacing $\rho$ with $r$, the Reissner-Nordström metric in isotropic
form is
\begin{align}
ds^{2} & =-\frac{\left(1-\frac{r_{\mathrm{S}}^{2}}{16r^{2}}+\frac{r_{\mathrm{Q}}^{2}}{4r^{2}}\right)^{2}}{\left[\left(1+\frac{r_{\mathrm{S}}}{4r}\right)^{2}-\frac{r_{\mathrm{Q}}^{2}}{4r^{2}}\right]^{2}}c_{0}^{2}dt^{2}+\left[\left(1+\frac{r_{\mathrm{S}}}{4r}\right)^{2}-\frac{r_{\mathrm{Q}}^{2}}{4r^{2}}\right]^{2}\left[dr^{2}+r^{2}\left(d\theta^{2}+\sin^{2}\left(\theta\right)d\varphi^{2}\right)\right]\nonumber \\
 & =-\left(1-\frac{r_{\mathrm{S}}}{r}+\frac{r_{\mathrm{Q}}^{2}}{r^{2}}\right)c_{0}^{2}dt^{2}+\left(1+\frac{r_{\mathrm{S}}}{r}-\frac{r_{\mathrm{Q}}^{2}}{2r^{2}}\right)\left[dr^{2}+r^{2}\left(d\theta^{2}+\sin^{2}\left(\theta\right)d\varphi^{2}\right)\right]+\ldots
\end{align}
which agrees to leading order with Eq. (\ref{eq:Static_Charged_Body});
but for the latter the electromagnetic energy only contributes to
the timelike part of the metric.

\subsection{Rotating Uncharged Axisymmetric Body\label{subsec:Rotating-Uncharged-Axisymmetric}}

The potential for an axisymmetric body is independent of $\varphi$,
i.e., depends only on $\left[r,\theta\right]$
\begin{equation}
\phi\equiv\phi\left(r,\theta\right).
\end{equation}
The metric for a rotating frame
\begin{equation}
\varphi\rightarrow\varphi\pm\omega t,\qquad d\varphi\rightarrow d\varphi\pm\frac{\omega}{c_{0}}c_{0}dt
\end{equation}
is
\begin{align}
ds^{2}= & -c_{0}^{2}dt^{2}+dr^{2}+r^{2}\left[d\theta^{2}+\sin^{2}\left(\theta\right)\left(d\varphi\pm\frac{\omega}{c_{0}}c_{0}dt\right)^{2}\right]\nonumber \\
= & -\left(1-\frac{\omega^{2}r^{2}}{c_{0}^{2}}\sin^{2}\left(\theta\right)\right)c_{0}^{2}dt^{2}\pm\frac{2\omega r^{2}}{c_{0}}\sin^{2}\left(\theta\right)d\varphi c_{0}dt+dr^{2}+r^{2}\left(d\theta^{2}+\sin^{2}\left(\theta\right)d\varphi^{2}\right)\label{eq:Rotating_Frame}
\end{align}
Similarly, for a central mass $M$ rotating with $\omega$ around
the $z$-axis, in the co-moving frame a local observer at rest in
the plane at distance $r$ observes a mass with angular momentum
\begin{equation}
L_{z}=M\omega r^{2}
\end{equation}
and the metric is given by Eq. (\ref{eq:Static_Charged_Body})
\begin{align}
ds^{2}= & -e^{\nicefrac{\phi_{\mathrm{m}}}{c_{0}^{2}}}c_{0}^{2}dt^{2}+e^{-\nicefrac{\phi_{\mathrm{m}}}{c_{0}^{2}}}\left\{ dr^{2}+r^{2}\left[d\theta^{2}+\sin^{2}\left(\theta\right)\left(d\varphi-\frac{a}{r^{2}c_{0}}c_{0}dt\right)^{2}\right]\right\} \nonumber \\
= & -\left(e^{\nicefrac{\phi_{\mathrm{m}}}{c_{0}^{2}}}-\frac{a^{2}}{r^{2}c_{0}^{2}}e^{-\nicefrac{\phi_{\mathrm{m}}}{c_{0}^{2}}}\sin^{2}\left(\theta\right)\right)c_{0}^{2}dt^{2}-\frac{2a}{c_{0}}e^{-\nicefrac{\phi_{\mathrm{m}}}{c_{0}^{2}}}\sin^{2}\left(\theta\right)d\varphi c_{0}dt\nonumber \\
 & +e^{-\nicefrac{\phi_{\mathrm{m}}}{c_{0}^{2}}}\left[dr^{2}+r^{2}\left(d\theta^{2}+\sin^{2}\left(\theta\right)d\varphi^{2}\right)\right]
\end{align}
where
\begin{equation}
a\equiv\frac{\omega r^{2}}{c_{0}}=\frac{L_{z}}{Mc_{0}}.
\end{equation}
By using Eq. (\ref{eq:Rotating_Frame}) to transform back to the rest
frame for the central mass one obtains
\begin{align}
ds^{2}= & -\left[e^{-\frac{r_{\mathrm{S}}}{r}}-\frac{a^{2}}{r^{2}}\left(e^{\frac{r_{\mathrm{S}}}{r}}-1\right)\sin^{2}\left(\theta\right)\right]c_{0}^{2}dt^{2}-2a\left(e^{\frac{r_{\mathrm{S}}}{r}}-1\right)\sin^{2}\left(\theta\right)d\varphi c_{0}dt\nonumber \\
 & +e^{\frac{r_{\mathrm{S}}}{r}}\left[dr^{2}+r^{2}\left(d\theta^{2}+\sin^{2}\left(\theta\right)d\varphi^{2}\right)\right]\nonumber \\
= & -\left(1-\frac{r_{\mathrm{S}}}{r}\right)c_{0}^{2}dt^{2}-\frac{2ar_{\mathrm{S}}}{r}\sin^{2}\left(\theta\right)d\varphi c_{0}dt+\left(1+\frac{r_{\mathrm{S}}}{r}\right)\left[dr^{2}+r^{2}\left(d\theta^{2}+\sin^{2}\left(\theta\right)d\varphi^{2}\right)\right]+\ldots\label{eq:Rot_Body}
\end{align}
where
\begin{equation}
r_{\mathrm{S}}\equiv\frac{2GM}{c_{0}^{2}}.
\end{equation}
Because the Schwarzschild solution for GR is for a spherically symmetric
body $\phi\equiv\phi\left(r\right)$, the solution for an axisymmetric
body $\phi\equiv\phi\left(r,\theta\right)$ can not be obtained by
transforming to a rotating system. The solution is the Kerr metric
\cite{Kerr} (Boyer-Lindquist form \cite{Boyer-Lindquist})
\begin{align}
ds^{2} & =-\left(1-\frac{r_{\mathrm{S}}r}{\varSigma}\right)c_{0}^{2}dt^{2}-\frac{2ar_{\mathrm{S}}r}{\varSigma}\sin^{2}\left(\theta\right)d\varphi c_{0}dt+\frac{\varSigma}{\varDelta}dr^{2}+\varSigma d\theta^{2}+\left(r^{2}+a^{2}+\frac{a^{2}r_{\mathrm{S}}r}{\rho^{2}}\right)\sin^{2}\left(\theta\right)d\varphi^{2}\nonumber \\
 & =-\left(1-\frac{r_{\mathrm{S}}}{r}\right)c_{0}^{2}dt^{2}-\frac{2ar_{\mathrm{S}}}{r}\sin^{2}\left(\theta\right)d\varphi c_{0}dt+\left(1+\frac{r_{\mathrm{S}}}{r}\right)dr^{2}+r^{2}\left(d\theta^{2}+\sin^{2}\left(\theta\right)d\varphi^{2}\right)+\ldots
\end{align}
where
\begin{equation}
\varSigma\equiv r^{2}\left(1+\frac{a^{2}}{r^{2}}\cos^{2}\left(\theta\right)\right),\qquad\varDelta\equiv r^{2}\left(1-\frac{r_{\mathrm{S}}}{r}+\frac{a^{2}}{r^{2}}\right).
\end{equation}
It can be written in quasi-isotropic coordinates by introducing \cite{Kerr-Newman_Isotropic_2}
\begin{equation}
r\rightarrow r\left(1+\frac{r_{\mathrm{S}}+2a}{4r}\right)\left(1+\frac{r_{\mathrm{S}}-2a}{4r}\right)
\end{equation}
which leads to
\begin{align}
ds^{2} & =-\alpha^{2}c_{0}^{2}dt^{2}+\psi^{4}\left[e^{\nicefrac{2\mu}{3}}\left(dr^{2}+r^{2}d\theta^{2}\right)+r^{2}\sin^{2}\left(\theta\right)e^{-\nicefrac{4\mu}{3}}\left(d\varphi+\beta^{\phi}dt\right)^{2}\right]\nonumber \\
 & =-\left(1-\frac{r_{\mathrm{S}}}{r}\right)c_{0}^{2}dt^{2}-\frac{2ar_{\mathrm{S}}}{r}\sin^{2}\left(\theta\right)d\varphi c_{0}dt+\left(1+\frac{r_{\mathrm{S}}}{r}\right)\left[dr^{2}+r^{2}\left(d\theta^{2}+\sin^{2}\left(\theta\right)d\varphi^{2}\right)\right]+\ldots
\end{align}
where
\begin{align}
\alpha^{2} & =\frac{\rho^{2}\varDelta}{\left(r^{2}+a^{2}\right)^{2}-\varDelta a^{2}\sin^{2}\left(\theta\right)}, & \beta^{\phi} & =-a\frac{r^{2}+a^{2}-\varDelta}{\left(r^{2}+a^{2}\right)^{2}-\varDelta a^{2}\sin^{2}\left(\theta\right)},\nonumber \\
\psi^{4} & =\frac{\rho^{\nicefrac{2}{3}}\left[\left(r^{2}+a^{2}\right)^{2}-\varDelta a^{2}\sin^{2}\left(\theta\right)\right]^{\nicefrac{1}{3}}}{r^{2}}, & e^{2\mu} & =\frac{\rho^{4}}{\left(r^{2}+a^{2}\right)^{2}-\varDelta a^{2}\sin^{2}\left(\theta\right)}
\end{align}
which agrees to leading order with Eq. (\ref{eq:Rot_Body}).

\subsection{Rotating Charged Axisymmetric Body}

The metric Eq. (\ref{eq:Metric_UFT_1}) is
\begin{equation}
ds^{2}=\left(g_{rs}-\frac{\kappa\varepsilon_{0}}{2}A_{r}A_{s}\right)dx^{r}dx^{s}
\end{equation}
with the vector potential\footnote{Retarded potentials are not required for the stationary case.}
\begin{equation}
A^{i}=\frac{\mu_{0}c_{0}}{4\pi}\int_{V}\frac{\rho u^{i}}{r}d^{3}x.
\end{equation}
For a moving point charge
\begin{equation}
A^{i}=\left[\frac{\phi_{\mathrm{q}}}{c_{0}},\bar{A}\right]=\frac{Q}{4\pi\varepsilon_{0}c_{0}r}\left[1,\bar{v}\right],\qquad A_{i}=g_{ir}A^{r}=\frac{Q}{4\pi\varepsilon_{0}r}\left[c_{0},\frac{\bar{v}}{c_{0}}\right]
\end{equation}
and the rotating frame metric in section \ref{subsec:Rotating-Uncharged-Axisymmetric}
\begin{equation}
A_{r}dx^{r}=\frac{Q}{4\pi\varepsilon_{0}r}\left(c_{0}dt+\frac{v_{\varphi}}{c_{0}}r\sin\left(\theta\right)d\varphi\right)=\frac{Q}{4\pi\varepsilon_{0}r}\left(c_{0}dt-\frac{\omega r^{2}\sin^{2}\left(\theta\right)}{c_{0}}d\varphi\right)=\frac{Q}{4\pi\varepsilon_{0}r}\left(c_{0}dt-a\sin^{2}\left(\theta\right)d\varphi\right)
\end{equation}
one obtains
\begin{equation}
-\frac{\kappa\varepsilon_{0}}{2}A_{r}A_{s}dx^{r}dx^{s}=-\frac{r_{\mathrm{Q}}^{2}}{r^{2}}\left(c_{0}^{2}dt^{2}-2a\sin^{2}\left(\theta\right)d\varphi c_{0}dt+a^{2}\sin^{4}\left(\theta\right)d\varphi^{2}\right)
\end{equation}
which when added to the metric for a static charged body in section
\ref{subsec:Static_Charged_Body} gives
\begin{align}
ds^{2}= & -\left[e^{-\frac{r_{\mathrm{S}}}{r}}+\frac{r_{\mathrm{Q}}^{2}-a^{2}\left(e^{\frac{r_{\mathrm{S}}}{r}}-1\right)\sin^{2}\left(\theta\right)}{r^{2}}\right]c_{0}^{2}dt^{2}-2a\left(e^{\frac{r_{\mathrm{S}}}{r}}-1-\frac{r_{\mathrm{Q}}^{2}}{r^{2}}\right)\sin^{2}\left(\theta\right)d\varphi c_{0}dt\nonumber \\
 & +e^{\frac{r_{\mathrm{S}}}{r}}\left\{ dr^{2}+r^{2}\left[d\theta^{2}+\left(1-\frac{r_{\mathrm{Q}}^{2}a^{2}\sin^{2}\left(\theta\right)}{r^{4}}\right)\sin^{2}\left(\theta\right)d\varphi^{2}\right]\right\} \nonumber \\
= & -\left(1-\frac{r_{\mathrm{S}}}{r}+\frac{r_{\mathrm{Q}}^{2}}{r^{2}}\right)c_{0}^{2}dt^{2}-2a\left(\frac{r_{\mathrm{S}}}{r}-\frac{r_{\mathrm{Q}}^{2}}{r^{2}}\right)\sin^{2}\left(\theta\right)d\varphi c_{0}dt\nonumber \\
 & +\left(1+\frac{r_{\mathrm{S}}}{r}\right)\left[dr^{2}+r^{2}\left(d\theta^{2}+\sin^{2}\left(\theta\right)d\varphi^{2}\right)\right]+\ldots\label{eq:Rot_Chg_Body}
\end{align}
where
\begin{equation}
a=\frac{L_{z}}{Mc_{0}},\qquad r_{\mathrm{S}}\equiv\frac{2GM}{c_{0}^{2}},\qquad r_{\mathrm{Q}}^{2}\equiv\frac{GQ^{2}}{4\pi\varepsilon_{0}c_{0}^{4}}.
\end{equation}
The electromagnetic field tensor is
\begin{equation}
F_{ij}=\partial_{i}A_{j}-\partial_{j}A_{i}=\begin{bmatrix}0 & \frac{Q}{4\pi\varepsilon_{0}r^{2}} & 0 & 0\\
-\frac{Q}{4\pi\varepsilon_{0}r^{2}} & 0 & 0 & \frac{Qa\sin^{2}\left(\theta\right)}{4\pi\varepsilon_{0}r^{2}}\\
0 & 0 & 0 & -\frac{2Qa\sin\left(\theta\right)\cos\left(\theta\right)}{4\pi\varepsilon_{0}r}\\
0 & -\frac{Qa\sin^{2}\left(\theta\right)}{4\pi\varepsilon_{0}r^{2}} & \frac{2Qa\sin\left(\theta\right)\cos\left(\theta\right)}{4\pi\varepsilon_{0}r} & 0
\end{bmatrix}.
\end{equation}
The corresponding Kerr-Newman metric for GR is (Boyer-Lindquist form)
\cite{Newman_1,Newman_2,Newman_3}
\begin{align}
ds^{2}= & -\frac{\varDelta-\frac{a^{2}}{c_{0}^{2}}\sin^{2}\left(\theta\right)}{\rho^{2}}c_{0}^{2}dt^{2}+\frac{2a\left(\varDelta-r^{2}-a^{2}\right)}{\rho^{2}}\sin^{2}\left(\theta\right)d\varphi c_{0}dt\nonumber \\
 & +\frac{\rho^{2}}{\varDelta}dr^{2}+\rho^{2}d\theta^{2}-\frac{a^{2}\varDelta\sin^{2}\left(\theta\right)-r^{4}-2r^{2}a^{2}-a^{4}}{\rho^{2}}\sin^{2}\left(\theta\right)d\varphi^{2}\nonumber \\
= & -\left(1-\frac{r_{\mathrm{S}}}{r}+\frac{r_{\mathrm{Q}}^{2}}{r^{2}}\right)c_{0}^{2}dt^{2}-\frac{2ar_{\mathrm{S}}}{r}\sin^{2}\left(\theta\right)d\varphi c_{0}dt+\left(1+\frac{r_{\mathrm{S}}}{r}\right)\left[dr^{2}+r^{2}\left(d\theta^{2}+\sin^{2}\left(\theta\right)d\varphi^{2}\right)\right]+\ldots
\end{align}
where
\begin{equation}
\rho^{2}=r^{2}\left(1+\frac{a^{2}}{r^{2}}\cos^{2}\left(\theta\right)\right),\qquad\varDelta=r^{2}\left(1-\frac{r_{\mathrm{S}}}{r}+\frac{a^{2}+r_{\mathrm{Q}}^{2}}{r^{2}}\right).
\end{equation}
As for the Kerr metric, it can be written in quasi-isotropic coordinates
by introducing \cite{Kerr-Newman_Isotropic_2}
\begin{equation}
r\rightarrow r\left(1+\frac{\frac{r_{\mathrm{S}}}{2}+\sqrt{a^{2}+r_{\mathrm{Q}}^{2}}}{2r}\right)\left(1+\frac{\frac{r_{\mathrm{S}}}{2}-\sqrt{a^{2}+r_{\mathrm{Q}}^{2}}}{2r}\right)
\end{equation}
which gives
\begin{align}
ds^{2}= & -\alpha^{2}c_{0}^{2}dt^{2}+\psi^{4}\left[e^{\nicefrac{2\mu}{3}}\left(dr^{2}+r^{2}d\theta^{2}\right)+r^{2}\sin^{2}\left(\theta\right)e^{-\nicefrac{4\mu}{3}}\left(d\varphi+\beta^{\phi}dt\right)^{2}\right]\nonumber \\
= & -\left(1-\frac{r_{\mathrm{S}}}{r}+\frac{r_{\mathrm{Q}}^{2}}{r^{2}}\right)c_{0}^{2}dt^{2}-\frac{2ar_{\mathrm{S}}}{r}\sin^{2}\left(\theta\right)d\varphi c_{0}dt\nonumber \\
 & +\left(1+\frac{r_{\mathrm{S}}}{r}-\frac{r_{\mathrm{Q}}^{2}}{r^{2}}\right)\left[dr^{2}+r^{2}\left(d\theta^{2}+\sin^{2}\left(\theta\right)d\varphi^{2}\right)\right]+\ldots
\end{align}
where
\begin{align}
\alpha^{2} & =\frac{\rho^{2}\varDelta}{\left(r^{2}+a^{2}\right)^{2}-\varDelta a^{2}\sin^{2}\left(\theta\right)}, & \beta^{\phi} & =-a\frac{r^{2}+a^{2}-\varDelta}{\left(r^{2}+a^{2}\right)^{2}-\varDelta a^{2}\sin^{2}\left(\theta\right)},\nonumber \\
\psi^{4} & =\frac{\rho^{\nicefrac{2}{3}}\left[\left(r^{2}+a^{2}\right)^{2}-\varDelta a^{2}\sin^{2}\left(\theta\right)\right]^{\nicefrac{1}{3}}}{r^{2}}, & e^{2\mu} & =\frac{\rho^{4}}{\left(r^{2}+a^{2}\right)^{2}-\varDelta a^{2}\sin^{2}\left(\theta\right)}.
\end{align}
The electromagnetic potential is
\begin{equation}
A_{r}dx^{r}=\frac{Qr}{4\pi\varepsilon_{0}\rho^{2}}\left(c_{0}dt-a\sin^{2}\left(\theta\right)d\varphi\right)
\end{equation}
with the field tensor
\begin{align}
F_{ij} & =\begin{bmatrix}0 & \frac{Q\left(r^{2}-a^{2}\cos^{2}\left(\theta\right)\right)}{4\pi\varepsilon_{0}\rho^{4}} & -\frac{2Qa^{2}r\sin\left(\theta\right)\cos\left(\theta\right)}{4\pi\varepsilon_{0}\rho^{4}} & 0\\
-\frac{Q\left(r^{2}-a^{2}\cos^{2}\left(\theta\right)\right)}{4\pi\varepsilon_{0}\rho^{4}} & 0 & 0 & \frac{Qa\left(r^{2}-a^{2}\cos^{2}\left(\theta\right)\right)\sin^{2}\left(\theta\right)}{4\pi\varepsilon_{0}\rho^{4}}\\
\frac{2Qra^{2}\sin\left(\theta\right)}{4\pi\varepsilon_{0}\rho^{4}} & 0 & 0 & -\frac{2Qar\left(a^{2}+r^{2}\right)\sin\left(\theta\right)\cos\left(\theta\right)}{4\pi\varepsilon_{0}\rho^{4}}\\
0 & -\frac{Qa\left(r^{2}-a^{2}\cos^{2}\left(\theta\right)\right)\sin^{2}\left(\theta\right)}{4\pi\varepsilon_{0}\rho^{4}} & \frac{2Qa\left(a^{2}+r^{2}\right)\sin\left(\theta\right)\cos\left(\theta\right)}{4\pi\varepsilon_{0}\rho^{4}} & 0
\end{bmatrix}\nonumber \\
 & =\begin{bmatrix}0 & \frac{Q}{4\pi\varepsilon_{0}r^{2}} & 0 & 0\\
-\frac{Q}{4\pi\varepsilon_{0}r^{2}} & 0 & 0 & \frac{Qa\sin^{2}\left(\theta\right)}{4\pi\varepsilon_{0}r^{2}}\\
0 & 0 & 0 & -\frac{2Qa\sin\left(\theta\right)\cos\left(\theta\right)}{4\pi\varepsilon_{0}r}\\
0 & -\frac{Qa\sin^{2}\left(\theta\right)}{4\pi\varepsilon_{0}r^{2}} & \frac{2Qa\sin\left(\theta\right)\cos\left(\theta\right)}{4\pi\varepsilon_{0}r} & 0
\end{bmatrix}+\ldots
\end{align}
The Kerr-Newman solution agrees to leading order with Eq. (\ref{eq:Rot_Chg_Body});
but for the latter, as in section \ref{subsec:Static_Charged_Body},
the electromagnetic energy does not contribute to the spacelike component
of the metric.

\section{Conclusion}

\noindent In 1945 Einstein concluded that \cite{Meaning_of_Rel.}:
\begin{quote}
\textquotedblleft \textit{The present theory of relativity is based
on a division of physical reality into a metric field (gravitation)
on the one hand, and into an electromagnetic field and matter on the
other hand. In reality space will probably be of a uniform character
and the present theory be valid only as a limiting case. For large
densities of field and of matter, the field equations and even the
field variables which enter into them will have no real significance.}\textquotedblright{}
\end{quote}
\noindent The dichotomy can be resolved by introducing a complex Randers
metric with a real valued scalar (gravitational) and complex valued
vector (electromagnetic) field for which Geometric Algebra and Lagrangian
mechanics provides a unified mathematical framework for: Electromagnetism,
Electrodynamics, Special and General Relativity (GR), and Gravitation.
The theory's predictions agree with GR; to leading order in the gravitational
constant. So the experimental results validate the presented framework
\& theory as well. Besides, because the metric is a functional of
the scalar and vector fields, solving for the latter provides the
metric solution directly. Hence, it is straightforward to obtain the
corresponding known GR metric solutions for a:
\begin{itemize}
\item spherically symmetric body (Schwarzschild),
\item static charged spherically symmetric body (Reissner-Nordström),
\item rotating uncharged axisymmetric body (Kerr),
\item rotating charged axisymmetric body (Kerr-Newman).
\end{itemize}
which agree to leading order in the gravitational constant; and are
free of spurious singularities. However, the electromagnetic energy
only contributes to the timelike component of the metric.

\section*{Acknowledgement}

\addcontentsline{toc}{section}{\textbf{Acknowledgement}}The author
became fascinated by Nordström's work 1983 after completing a class
in General Relativity by self-studies at the Dept. of Theor. Elementary
Particle Physics, Lund Univ. Ten years later, 1995, a semester was
spent at Prof. (now Emeritus) Olavi Nevanlinna's Dept. of Mathematics
and Systems Analysis, Helsinki Univ. of Tech., now Aalto Univ., pursuing
post grad self-studies. Coincidently, Nordström had been Prof. there
\cite{Isaksson}. Similarly, my thanks to a colleague Prof. Vladimir
Litvinenko, Stony Brook Univ. and Brookhaven Natl. Lab. Simply put,
for their: intellectual curiosity, honesty, and integrity. Also, my
thanks to the people behind the ArXiv Server and the Expert(s) who
scrutinized the first pre-print.


\begin{thebibliography}{10}
\bibitem{Meaning_of_Rel.}\addcontentsline{toc}{section}{\textbf{References}}A.\ Einstein
``The Meaning of Relativity'' (MJF Books, 1997) \href{https://books.google.com/books?id=u6bgAAAAMAAJ}{ISBN 1567311369}.

\bibitem{Bengtsson}J.\ Bengtsson ``General Relativity Revisited:
Generalized Nordström Theory'' \href{https://arxiv.org/abs/1610.03702}{arXiv:1610.03702 (2016)}.

\bibitem{Weyl_1}H.\ Weyl ``Gravitation und Elektrizität'' \href{http://articles.adsabs.harvard.edu/cgi-bin/get_file?pdfs/SPAW./1918/1918SPAW.......465W.pdf}{Sitzungsber. Preuss. Akad. Wissensch. Berlin 465\textendash 480 (1918)}.

\bibitem{Weyl_2}H.\ Weyl ``Raum. Zeit. Materie'' (\href{https://books.google.com/books?id=6jNPAQAAMAAJ}{Springer, 1918})
transl. H.L.\ Brose ``Space, Time, Matter'' (Dover, 1952) \href{https://books.google.com/books?id=GdwPAQAAMAAJ}{ISBN 0486602672}.

\bibitem{Einstein_1918}A.\ Einstein ``Nachtrag'' \href{http://articles.adsabs.harvard.edu/cgi-bin/get_file?pdfs/SPAW./1918/1918SPAW.......465W.pdf}{Sitzungsber. Preuss. Akad. Wissensch. Berlin, 478-478 (1918)}.

\bibitem{London}F.\ London ``Die Theorie von Weyl und die Quantenmechanik''
\href{http://dx.doi.org/10.1007/BF01505037}{F. Naturwissenschaften 15, 187-187 (1927)}.

\bibitem{Randers}G.\ Randers ``On an Asymmetrical Metric in the
Four-Space of General Relativity'' \href{http://dx.doi.org/10.1103/PhysRev.59.195}{Phys. Rev. 59, 195-199 (1941)}.

\bibitem{Soh}H.P.\ Soh ``Theory of Gravitation and Electromagnetism''
\href{https://dx.doi.org/10.1002/sapm1933121298}{J. Math. Phys. 12, 298-305 (1933)}.

\bibitem{Einstein_Complex_1}A.\ Einstein ``A Generalization of
the Relativistic Theory of Gravitation'' \href{ http://www.jstor.org/stable/1969197}{Ann. Math. 46, 578-584 (1945)}.

\bibitem{Einstein_Complex_2}A.\ Einstein ``A Generalization of
the Relativistic Theory of Gravitation, II'' \href{http://dx.doi.org/10.2307/1969231}{Ann. Math. 47, 731-741 (1946)}.

\bibitem{Einstein_Complex_3}A.\ Einstein ``A Generalized Theory
of Gravitation'' \href{https://doi.org/10.1103/RevModPhys.20.35}{Rev. Mod. Phys. 20, 35-39 (1948)}.

\bibitem{Schwarzschild}K.\ Schwarzschild \textquotedblleft Über
das Gravitationsfeld eines Massenpunktes nach der Einsteinschen Theorie\textquotedblright{}
\href{http://articles.adsabs.harvard.edu/cgi-bin/get_file?pdfs/SPAW./1916/1916SPAW.......189S.pdf}{Sitzungsberichte der Königlich Preussischen Akademie der Wissenschaften 1, 189-196 (1916)}.

\bibitem{Einstein_GR}A.\  Einstein ``Die Grundlage der allgemeinen
Relativitätstheorie'' \href{http://dx.doi.org/10.1002/andp.19163540702}{Ann. Phys. 49 (7), 769-822 (1916)}.

\bibitem{Grassmann}H.\ Grassmann ``Die lineale Ausdehnungslehre,
ein neuer Zweig der Mathematik'' (\href{https://books.google.com/books?id=nyiF33VKDOUC}{O. Wigand, 1844}).

\bibitem{Cartan}É\ Cartan ``Sur certaines expressions différentielles
et le problème de Pfaff'' \href{https://doi.org/10.24033/asens.467}{Annales Scientifiques de l'École Normale Supérieure Série 3 16, 239-332 (Gauthier-Villars, Paris, 1899)}.

\bibitem{Cartan_Maxwell}É\ Cartan ``Sur les variétés à connexion
affine, et la théorie de la relativité généralisée'' \href{http://eudml.org/doc/81425}{Ann. Ec. Norm. Sup. 41, 1-25 (1924)}.

\bibitem{Hamilton}W.R.\ Hamilton ``On Quaternions'' \href{http://dx.doi.org/10.1080/14786444408644923}{Philos. Mag. 25, 10-13 (1844)}.

\bibitem{Peano}G.\ Peano ``Calcolo geometrico secondo l'Ausdehnungslehre
di H. Grassmann'' (\href{https://books.google.com/books?id=5LJi3dxLzuwC}{Fratelli Bocca, 1888})
transl. ``Geometric Calculus: According to the Ausdehnunglehre of
H. Grassmann'' L.C.\ Kannenberg (Birkhäuser, 2000) \href{https://books.google.com/books?id=wK4bloDOohEC}{ISBN 0817641262}.

\bibitem{Hestenes}D.\ Hestenes, G.\ Sobczyk ``Clifford Algebra
to Geometric Calculus: A Unified Language for Mathematics and Physics''
(Springer, 1987) \href{https://books.google.com/books?id=dScR5zwrheYC}{ISBN 9027725616}.

\bibitem{Hestenes_2}D.\ Hestenes, \textquotedblleft Spacetime Geometry
with Geometric Calculus\textquotedblright{} \href{https://www.researchgate.net/publication/228380906_Spacetime_Geometry_with_Geometric_Calculus}{7th Intl. Conf. Clifford Algebras and their Applications, Tolouse, France, 2005}.

\bibitem{Doran}C.\ Doran, A.\ Lasenby ``Geometric Algebra for
Physicists'' (Cambridge University Press, 2003) \href{https://books.google.com/books?id=VW4yt0WHdjoC}{ISBN 0521480221}.

\bibitem{Vold}T.G.\ Vold ``An Introduction to Geometric Calculus
and its Application to Electrodynamics'' \href{http://dx.doi.org/10.1119/1.17202}{Am. J. Phys. 61, 505-505 (1993)}.

\bibitem{Gibbs}J.W.\ Gibbs ``Elements of Vector Analysis Arranged
for the Use of Students in Physics'' (\href{https://openlibrary.org/books/OL23435675M}{Tuttle, Morehouse \&{} Taylor, 1884}).

\bibitem{Weber}H.M.\ Weber ``Die partiellen Differential-Gleichungen
der mathematischen Physik nach Riemann's Vorlesungen'' Zweiter Band
p. 348 (\href{https://archive.org/details/diepartiellendi00webegoog}{Friedrich Vieweg und Sohn, Braunscheweig, 1901}).

\bibitem{Silberstein}L.\ Silberstein ``Elektromagnetische Grundgleichungen
in bivectorieller Behandlung'' \href{https://doi.org/10.1002/andp.19073270313}{Ann. der Phys. 22, 579-586 (1907)}.

\bibitem{Christoffel}E.B.\ Christoffel ``Über die Bestimmung der
Gestalt einer krummen Oberfläche durch lokale Messungen auf derselben''
\href{http://gdz.sub.uni-goettingen.de/dms/load/img/?PID=GDZPPN002152533}{J. Reine Angew. Math. 64, 193-209 (1865)}.

\bibitem{Eddington}A.S.\ Eddington \textquotedblleft The Mathematical
Theory of Relativity\textquotedblright{} (\href{https://books.google.se/books?id=hy2CMQEACAAJ}{ISBN 0521091659 (University Press, 1924)}).

\bibitem{Reissner}H.\ Reissner. ``Über die Eigengravitation des
elektrischen Feldes nach der Einsteinschen Theorie'' \href{https://dx.doi.org/10.1002\%2Fandp.19163550905}{Ann. Phys. 50, 106\textendash 120 (1916)}.

\bibitem{Nordstr=0000F6m_3}G.\ Nordström ``On the Energy of the
Gravitational Field in Einstein\textquoteright s Theory\textquotedblright{}
\href{https://archive.org/stream/proceedingsofsec202koni\#page/1237/mode/2up}{Verhandl. Koninkl. Ned. Akad. Wetenschap. 20 (2), 1238\textendash 1245 (1918)}.

\bibitem{Kerr}R.P.\ Kerr ``Gravitational Field of a Spinning Mass
as an Example of Algebraically Special Metrics'' \href{https://doi.org/10.1103/PhysRevLett.11.237}{Phys. Rev.Lett. 11, 237\textendash 238 (1963)}.

\bibitem{Boyer-Lindquist}R.H.\ Boyer, W.R.\ Lindquist. ``Maximal
Analytic Extension of the Kerr Metric'' \href{https://doi.org/10.1063\%2F1.1705193}{J. Math. Phys. 8, 265-281 (1967)}.

\bibitem{Kerr-Newman_Isotropic_2}G.B.\ Cook ``Initial Data for
Numerical Relativity'' \href{https://arxiv.org/abs/gr-qc/0007085}{arXiv:gr-qc/0007085 (2000)}.

\bibitem{Newman_1}A.I.\ Janis, E.T.\ Newman ``Structure of Gravitational
Sources'' \href{http://dx.doi.org/10.1063/1.1704349}{J. Math. Phys. 6, 902-914 (1965)}.

\bibitem{Newman_2}E.\ Newman, J.\ Allen ``Note on the Kerr Spinning-Particle
Metric'' \href{http://dx.doi.org/10.1063/1.1704350}{J. Math. Phys. 6, 915\textendash 917 (1965)}.

\bibitem{Newman_3}E.T.\ Newman, E.\ Couch, K.\ Chinnapared, A.\ Exton,
A.\ Prakash, R.\ Torrence ``Metric of a Rotating, Charged Mass''
\href{http://dx.doi.org/10.1063/1.1704351}{J. Math. Phys. 6, 918-919 (1965)}.

\bibitem{Isaksson}E.\ Isaksson \textquotedblleft Gunnar Nordström
(1881 - 1923). On Gravitation and Relativity\textquotedblright{} \href{http://www.helsinki.fi/~eisaksso/nordstrom/nordstrom.html}{XVIIth Intl. Congr. Hist. Sci., Univ. of California, Berkeley, July 31 - August 8, 1985}.
\end{thebibliography}
\end{document}